%% file: CoST.tex
\documentclass[sigconf,authorversion]{acmart}
\usepackage{enumitem}
\usepackage{booktabs}
\usepackage{marvosym}
\usepackage{multirow}

\begin{document}

\copyrightyear{2024}
\acmYear{2024}
\setcopyright{acmlicensed}\acmConference[RecSys '24]{18th ACM Conference
on Recommender Systems}{October 14--18, 2024}{Bari, Italy}
\acmBooktitle{18th ACM Conference on Recommender Systems (RecSys '24),
October 14--18, 2024, Bari, Italy}
\acmDOI{10.1145/3640457.3688178}
\acmISBN{979-8-4007-0505-2/24/10}

\title[CoST: Contrastive Quantization based Semantic Tokenization for Generative Recommendation]{CoST: Contrastive Quantization based Semantic Tokenization \\for Generative Recommendation}


\author{Jieming Zhu$^*$}
\affiliation{%
 \institution{Huawei Noah's Ark Lab}
 \city{Shenzhen}
 \country{China}
}
\email{jiemingzhu@ieee.org}

\author{Mengqun Jin$^*$}
\affiliation{%
 \institution{Shenzhen International Graduate School, Tsinghua University}
 \city{Shenzhen}
 \country{China}
}
\email{jinmq22@mails.tsinghua.edu.cn}

\author{Qijiong Liu}
\affiliation{%
 \institution{The Hong Kong Polytechnic University}
 \city{Hong Kong}
 \country{China}
}
\email{liu@qijiong.work}

\author{Zexuan Qiu}
\affiliation{%
\institution{The Chinese University of Hong Kong}
\city{Hong Kong}
\country{China}
}
\email{qzexuan@link.cuhk.edu.hk}

\author{Zhenhua Dong}
\affiliation{%
 \institution{Huawei Noah's Ark Lab}
 \city{Shenzhen}
 \country{China}
}
\email{dongzhenhua@huawei.com}

\author{Xiu Li\textsuperscript{\Letter}}
\affiliation{%
 \institution{Shenzhen International Graduate
School, Tsinghua University}
 \city{Shenzhen}
 \country{China}
}
\email{li.xiu@sz.tsinghua.edu.cn}

\thanks{$^*$ \hspace{0.5ex}Equal contribution. The work was done during Mengqun's internship at Huawei.}
\thanks{\textsuperscript{\Letter} Corresponding author.}

\renewcommand{\authors}{Jieming Zhu, Mengqun Jin, Qijiong Liu, Zexuan Qiu, Zhenhua Dong, Xiu Li}
\renewcommand{\shortauthors}{Zhu et al.}

\input{Content/Abstract}

\ccsdesc[500]{Information systems~Recommender systems}
\ccsdesc[500]{Information systems~Personalization}

\keywords{Generative Recommendation, Semantic Tokenization; Contrastive Quantization}

\settopmatter{printacmref=true}


\maketitle

\input{Content/Introduction}
\input{Content/RelatedWork}

\input{Content/Approach}

\input{Content/Experiments}

\input{Content/Conclusion}

\balance
\bibliographystyle{ACM-Reference-Format}
\bibliography{CoST}

\end{document}

%% file: Content/Abstract.tex
\begin{abstract}

Embedding-based retrieval serves as a dominant approach to candidate item matching for industrial recommender systems. With the success of generative AI, generative retrieval has recently emerged as a new retrieval paradigm for recommendation, which casts item retrieval as a generation problem. Its model consists of two stages: semantic tokenization and autoregressive generation. The first stage involves item tokenization that constructs discrete semantic tokens to index items, while the second stage autoregressively generates semantic tokens of candidate items. Therefore, semantic tokenization serves as a crucial preliminary step for training generative recommendation models. Existing research usually employs a vector quantizier with reconstruction loss (e.g., RQ-VAE) to obtain semantic tokens of items, but this method fails to capture the essential neighborhood relationships that are vital for effective item modeling in recommender systems. In this paper, we propose a contrastive quantization-based semantic tokenization approach, named CoST, which harnesses both item relationships and semantic information to learn semantic tokens. Our experimental results highlight
the significant impact of semantic tokenization on generative recommendation performance, with CoST achieving up to a 43\% improvement in Recall@5 and 44\% improvement in NDCG@5 on the MIND dataset over previous baselines.

\end{abstract}

%% file: Content/Introduction.tex
\section{Introduction}

Industrial recommender systems typically employ a retrieve-then-rank pipeline to serve online requests while adhering to strict response time requirements. In this pipeline, embedding-based retrieval plays a crucial role by efficiently retrieving highly relevant items from a large candidate pool for the subsequent fine-grained ranking stage. However, current embedding-based retrieval approaches are limited by either two-tower models~\cite{covington2016deep,MNS,CBNS,EBR2} or graph networks~\cite{PinSage,EGES,LightSAGE,UltraGCN}, which rely on dot-product (or cosine) similarity for approximate nearest neighbor (ANN) search in a pre-computed vector store (e.g., Faiss~\cite{Faiss, Faiss2}). While ANN search systems facilitate efficient top-k item retrieval, they constrain model capacity by using simple dot-product similarity and lead to disjoint optimization of embedding models and ANN indexes. 

\begin{figure}[!t]
 \centering
 \includegraphics[width=\linewidth]{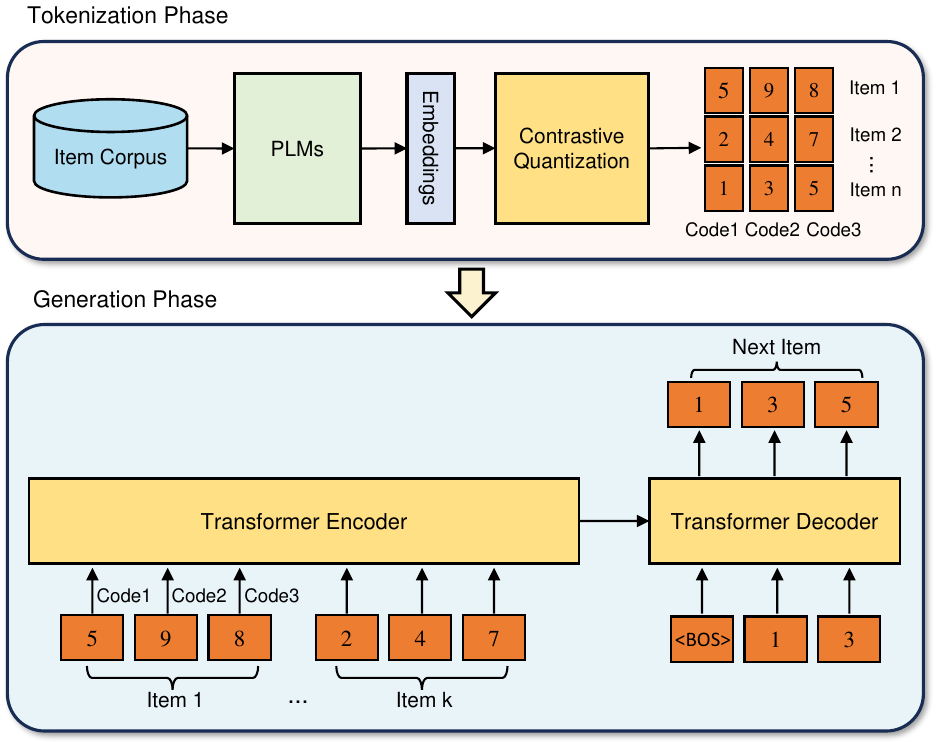}
 \caption{A framework of generative recommendation including tokenization phase and generation phase.}
 \label{fig:overview}
\end{figure}

With the recent success of generative AI techniques, generative retrieval~\cite{GR,tiger,EAGER} has emerged as a new retrieval paradigm for recommendation, redefining item retrieval as a sequence generation problem. TIGER~\cite{tiger} exemplifies this approach with its two-stage model: semantic tokenization and autoregressive generation. As illustrated in Figure~\ref{fig:overview}, in the tokenization stage, the model maps each item to a sequence of discrete semantic tokens (also known as semantic IDs or codes), aiming to capture and preserve item content semantics within these tokens. This ensures that semantically similar items are assigned similar tokens. In the generation stage, an encoder-decoder architecture is employed to encode historical item sequences and decode target item tokens autoregressively. During inference, by using beam search with token-to-item mapping, the model can directly generate top-k candidate item IDs. In this way, generative recommendation models like TIGER~\cite{tiger} and EAGER~\cite{EAGER} facilitate fine-grained token-level relevance matching between users and items while eliminating the need for non-differentiable, external ANN search systems.

In this process, semantic tokenization plays a crucial role as a preliminary step for training generative recommendation models, reducing the vocabulary size from millions of item IDs to thousands of tokens, which enables more efficient decoding without the need for ANN indexes. To learn high-quality semantic tokens, existing work typically leverages textual descriptions of items with pretrained language models (e.g., Sentence-T5~\cite{sentenceT5}) to derive their semantic embeddings. Subsequently, some techniques such as vector quantization~\cite{tiger,betterSemantic,VQSurvey} and hierarchical k-means clustering~\cite{tay2022transformer,recforest,EAGER} are applied to discretize semantic embeddings into semantic tokens (w.r.t. codes or clusters). Among them, RQ-VAE, first introduced in the TIGER framework~\cite{tiger}, stands out as a leading approach for semantic tokenization. RQ-VAE employs a residual quantization-based autoencoder, optimized through reconstruction loss. However, its reliance on the autoencoder architecture, which compresses and discretizes semantic embeddings for each item independently, can result in a suboptimal distribution of the learned semantic token space when applied to generative recommendation models. For example, due to the long-tailed item distribution observed in practice, RQ-VAE often produces identical token sequences for similar items within the densest regions, making them difficult to differentiate. This thus motivates us to capture essential neighborhood relationships among items that are vital for recommendation models.




To achieve this goal, this paper introduces CoST, a \underline{Co}ntrastive quantization-based \underline{S}emantic \underline{T}okenization method, which not only captures the semantic information of items but also integrates their neighborhood relationships. Building on the RQ-VAE structure, CoST encodes semantic item embeddings and then uses an iterative residual quantization process in the latent space to assign these embeddings to the closest codes (i.e., tokens) from a learnable codebook (i.e., vocabulary). A decoder further reconstructs the item embeddings from their corresponding semantic codes and codebook embeddings. However, unlike the reconstruction-based RQ-VAE, which aims to precisely reconstruct item embedding vectors using MSE loss, CoST adopts a contrastive learning framework to capture neighborhood relationships among items during quantization. Specifically, we enforce that reconstructed vectors are closer to their corresponding input vectors than to other vectors within a batch. This method thus preserves the neighborhood information between the input embedding and its reconstructed counterpart while enhancing dissimilarity from other items, facilitating clearer differentiation between each item and its neighbors. Our approach relaxes the requirement for exact reconstruction and instead prioritizes pairwise similarities between items. By leveraging contrastive loss, which maximizes the top-one probability within a batch, we more effectively capture the underlying distribution of item similarities. This provides a key advantage for recommendation tasks compared to RQ-VAE.

To evaluate the effectiveness of our approach, we use TIGER~\cite{tiger} as the baseline generative recommendation model and assess CoST as a replacement for RQ-VAE during the semantic tokenization phase. Our experiments on multiple benchmark datasets highlight the significant impact of semantic tokenization on generative recommendation performance. Notably, on the MIND dataset~\cite{MIND}, CoST achieves substantial improvements, with up to a 43\% increase in Recall@5 and a 44\% increase in NDCG@5 compared to the reconstruction-based RQ-VAE. These results underscore the pivotal role of semantic tokenization in generative recommendation and pave the way for future research in this direction.





%% file: Content/RelatedWork.tex
\section{Related Work}

\textbf{Generative AI for Recommendation.} With the recent success of generative AI, generative recommendation has emerged as a trending research topic. On one hand, researchers are exploring the application of sequence generation models for recommendation tasks. Among them, generative retrieval (GR) \cite{GR} stands out as a promising technique for retrieving relevant documents from a database by reframing the retrieval process as a sequence generation problem. Unlike traditional embedding-based retrieval methods, which rely on two-tower models for embedding computation and external ANN search systems for top-k lookups, GR generates document tokens sequentially, including document titles, document IDs, or assigned semantic IDs \cite{tay2022transformer}. Notably, GENRE \cite{genre} pioneered the use of a transformer architecture for entity retrieval, generating entity names token-by-token based on a given query. DSI \cite{tay2022transformer} and NCI introduced the concept of assigning structured semantic IDs to documents and training encoder-decoder models for generative document retrieval. Upon receiving a query, these models generate semantic IDs autoregressively to retrieve documents. Following this paradigm, TIGER \cite{tiger} and EAGER \cite{EAGER} were developed as generative item retrieval models for recommender systems. Additionally, there have been efforts to apply non-autoregressive generative models for tasks such as item list continuation \cite{FANS} and item list re-ranking \cite{NAGM}. On the other hand, some research explores leveraging the generation and reasoning capabilities of large language models (LLMs) for recommendation tasks. Examples include generating item embeddings \cite{KAR,NoteLLM}, item descriptions \cite{ItemLanguage}, conversational responses \cite{LLMCRS}, explanations \cite{LubosTFEL24}, or summaries \cite{GNR} to support recommendations.

\noindent\textbf{Semantic Indexing and Tokenization.} 
With the rising popularity of generative recommendation, how to index items in these models has become a key challenge. \cite{p5index} conducts in-depth research on item indexing methods for LLM-based recommendation models, such as P5 \cite{p5}. While simple indexing methods, such as random indexing and title indexing, are straightforward, they are not easily scalable to industrial-sized recommendation scenarios. To address this, research has focused on semantic indexing \cite{tay2022transformer,betterSemantic}, which aims to index items based on coherent content information. Two primary techniques are used, including vector quantization \cite{tiger,betterSemantic,VQSurvey} and hierarchical k-means clustering \cite{tay2022transformer,recforest,EAGER}. Given their discrete nature and similarities to visual tokenization \cite{VQVAE,VisualTokenizer}, this process is also referred to as semantic tokenization \cite{LiuH0ZK024}. Specifically, TIGER \cite{tiger} and LC-Rec \cite{LC-Rec} apply residual quantization (RQ-VAE) to textual embeddings derived from item titles and descriptions for tokenization. Recforest~\cite{recforest} and EAGER~\cite{EAGER} utilize hierarchical k-means clustering on item textual embeddings to obtain cluster indexes as tokens. Furthermore, recent studies such as EAGER \cite{EAGER}, TokenRec \cite{TokenRec}, and LETTER \cite{LETTER} explore integrating both semantic and collaborative information into the tokenization process. However, this work focuses on how to leverage contrastive quantization~\cite{ContrastiveQuant1,ContrastiveQuant2} to enhance semantic tokenization.

%% file: Content/Approach.tex
\section{Approach}
Figure~\ref{fig:overview} illustrates the generative recommendation framework, which comprises two phases: semantic tokenization and autoregressive generation. In this section, we first introduce our proposed CoST approach for semantic tokenization, followed by a description of its integration with the generation task.



\begin{figure}[t]
 \centering
 \includegraphics[width=\linewidth]{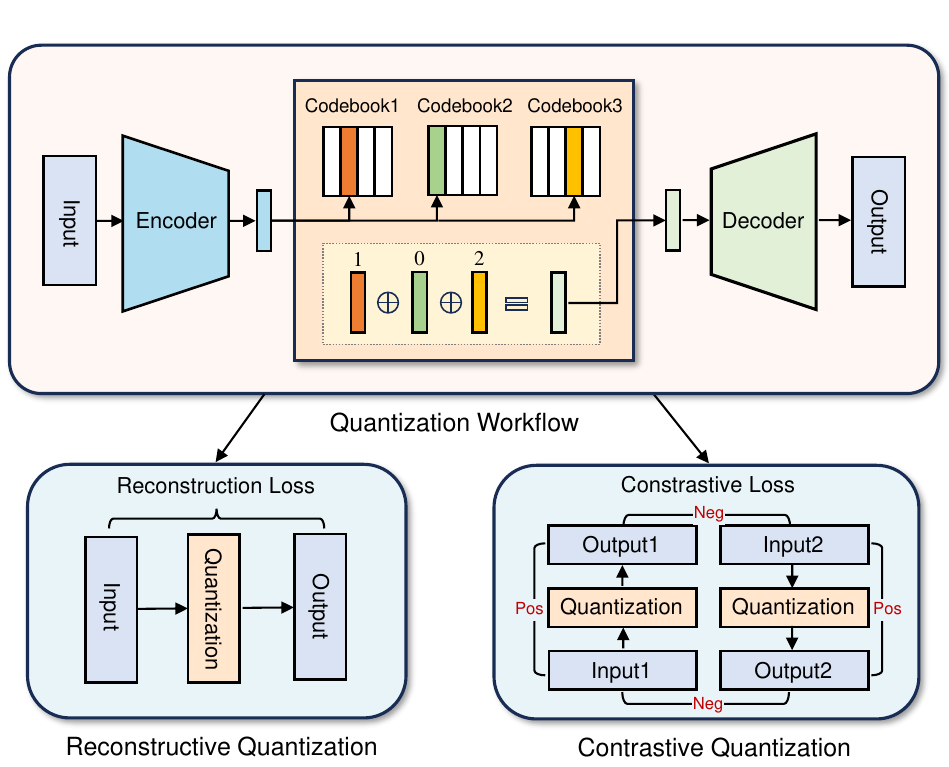}
 \caption{The vector quantization workflow trained via reconstructive quantization and contrastive quantization.}
 \label{fig:RQVAE}
 \vspace{-2ex}
\end{figure}

\subsection{Contrastive Quantization-based Semantic Tokenization}


As shown in Figure~\ref{fig:overview}, we initially employ pretrained text encoders, such as Sentence-T5 \cite{sentenceT5} and E5 \cite{E5}, to derive item embeddings from their textual content. For items with multiple textual fields, such as category, title, and description, we concatenate these fields using a prompt template, for example: "Item category: <category>. Title: <title>. Description: <description>." Once the item embeddings are obtained, we utilize a residual-quantized variational autoencoder (RQ-VAE) \cite{RQVAE_CV} structure to learn discrete tokens of items, a process illustrated in Figure~\ref{fig:RQVAE}.



Specifically, to tokenize an input vector $\mathbf{x}$, we employ an encoder $\mathrm{E}$ (e.g., MLP) to project $\mathbf{x}$ onto a low-dimensional latent representation $\mathbf{z} = \mathrm{E}(\mathbf{x})$. Residual quantization is an iterative process that quantizes the residual vector at each step, thereby generating a tuple of tokens. Initially, the residual is defined as $\mathbf{r}_0 = \mathbf{z}$. We then establish a set of codebooks $\mathbf{C}_i = \{\mathbf{e}^k_i | k=1,\ldots, K\}, \forall i \in [1, M]$, where $M$ represents the number of codebooks and $K$ denotes the size of each codebook. To approximate $\mathbf{r}_0$, we perform a nearest neighbor search using the first codebook $\mathbf{C}_1$, selecting the closest code vector with its index $c_1 = \arg \min_k \|\mathbf{r}_0 - \mathbf{e}^k_1\|_2$. 
Next, we compute the residual at the second level, $\mathbf{r}_1 = \mathbf{r}_0 - \mathbf{e}_1^{c_1}$, which captures the information not represented by the first code vector. This residual is then approximated using the second codebook $\mathbf{C}_2$. By iteratively repeating this procedure $M$ times, we generate a tuple of $M$ codes $(c_1,\dots,c_M)$, which are denoted as semantic tokens.


\subsubsection{Reconstructive Quantization}

As illustrated in the lower left of Figure~\ref{fig:RQVAE}, existing work typically trains the quantizer by minimizing the reconstruction loss between the reconstructed vector and the input vector. Specifically, given the semantic tokens after quantization, we construct an approximated vector $\widehat{\mathbf{z}}$ by summing over $M$ code vectors: $\widehat{\mathbf{z}} = \sum_{i=1}^{M} \mathbf{e}_i^{c_i}$. This approximated vector $\widehat{\mathbf{z}}$ is then fed into a decoder (e.g., an MLP) to produce $\widehat{x}$, which serves as a reconstruction of the input $\mathbf{x}$. Similar to the popular VQ-VAE technique~\cite{VQVAE}, the training loss function for reconstructive quantization is defined as:
\begin{equation}
 \mathcal{L}_\text{re} = \mathcal{L}_\text{mse} + \mathcal{L}_\text{rq}
\end{equation}
which consists of two main terms: $\mathcal{L}_\text{mse}$, which represents the MSE reconstruction loss and $\mathcal{L}_\text{rq}$, which denotes the quantization loss for codebook learning.
\begin{eqnarray}
 \mathcal{L}_\text{mse} &=& \|\mathbf{x}- \widehat{\mathbf{x}}\|_2^2  \\
 \mathcal{L}_\text{rq} &=& \sum\nolimits_{i=1}^M{\|\mathrm{sg}(\mathbf{r}_{i-1})- \mathbf{e}_i^{c_i}\|_2^2} + \beta\|\mathbf{r}_{i-1}- \mathrm{sg}(\mathbf{e}_i^{c_i})\|_2^2
\end{eqnarray}
where $\mathrm{sg}$ denotes the stop-gradient operation~\cite{stopgradient}. The loss term $\mathcal{L}_\text{rq}$ acts as a straight-through estimator for quantization, enabling end-to-end training of the encoder, decoder, and codebooks.

To prevent the codebook collapse issue \cite{CodeCollapse}, which refers to the under-utilization of codebook vectors, we employ an initialization approach based on k-means clustering \cite{RQVAE_sound}. During the initial training batch, the k-means algorithm is applied, and the resulting centroids are used to initialize the codebook vectors.


\subsubsection{Contrastive Quantization}

Reconstructive quantization aims for the precise reconstruction of each item embedding independently, which can lead to a suboptimal distribution of the learned semantic token space when applied to generative recommendation models. Unlike visual tokens, which require pixel-level decoding for image generation \cite{VQVAE}, generative retrieval tasks focus more on differentiating among items. Consequently, we introduce a contrastive quantization-based semantic tokenization method, CoST, to capture both semantic information of items and their neighborhood relationships.

Specifically, as shown in the lower right of Figure~\ref{fig:RQVAE}, we enforce the positive pair between the input vector $\mathbf{x}_0$ and the reconstructed vector $\widehat{\mathbf{x}_0}$ to be closer than other vectors $\{\widehat{\mathbf{x}_i} | i=1, \ldots, K\}$ within a batch. This is achieved using the following contrastive loss:
\begin{equation}
 \mathcal{L}_{\text{cl}} = - \log \frac{\exp{\big(\langle \mathbf{x}_0, \widehat{\mathbf{x}_0}\rangle /\tau}\big)}{\sum_{i=0}^K \exp{\big(\langle \mathbf{x}_0, \widehat{\mathbf{x}_i}
 \rangle /\tau}\big)}~~,
\end{equation}
where $\langle \cdot,\cdot\rangle$ denotes the cosine similarity between two vectors, and $\tau$ represents the temperature parameter that regulates the level of concentration. The $\mathcal{L}_{\text{cl}}$ loss relaxes the requirement for exact reconstruction and instead prioritizes pairwise similarities between items, thereby preserving the neighborhood information between the input embedding and its reconstructed counterpart while enhancing the dissimilarity from other items. This facilitates clearer differentiation between each item and the others. Finally, the loss function for contrastive quantization is defined as:
\begin{equation}
 \mathcal{L}_\text{co} = \alpha \mathcal{L}_\text{cl} + \mathcal{L}_\text{rq}~~,
\end{equation}
where $\alpha$ is a hyperparameter used to balance the two terms.


\begin{table}[t]
\centering
\caption{Dataset statistics.}
\label{dataset}
\setlength{\tabcolsep}{2mm}{
\begin{tabular}{c|ccccc}
\toprule[1pt]
Datasets & \#User & \#Item & Seq\_Len & Average & Median \\ \midrule
MIND & 29,207 & 12,251 & [15, 70] & 25.06 &22 \\
Office(S) & 2,868 & 14,618 & [10, 20] & 13.21 & 12 \\
Office(L) &16,696 &37,347	& [5, 50] &8.38 &12 \\ \bottomrule[1pt]
\end{tabular}}
\vspace{-1ex}
\end{table}

\subsection{Autoregressive Generation}

Following the TIGER model~\cite{tiger}, we employ an encoder-decoder transformer architecture for generative recommendation. Specifically, we tokenize each item into a tuple of tokens, resulting in a behavior sequence $\mathbf{s}=(\mathbf{c}_{1,1},\dots,\mathbf{c}_{1,M},\mathbf{c}_{2,1},\dots, \mathbf{c}_{2,M},\dots,\mathbf{c}_{N,1},\dots,\mathbf{c}_{N,M})$, where $N$ represents the sequence length of items. The model is then trained as a seq2seq learning task for autoregressive token generation. Given the token sequence of the first $i-1$ items, the model predicts the tokens $(\mathbf{c}_{i,1},\dots,\mathbf{x}_{i,M})$ for the $i$-th item $t_i$ autoregressively, rather than generating an item ID directly. After training, we use beam search to generate tokens and then lookup the token-item table to obtain the corresponding items for recommendation.


%% file: Content/Experiments.tex
\section{Experiments}
\subsection{Experimental Settings}

\noindent{\bfseries Datasets.} 
We conducted experiments on two real-world datasets: MIND \cite{MIND}, and Amazon Office. The statistics of the preprocessed datasets are summarized in Table~\ref{dataset}, including the range (Seq\_Len), average, and median of behavior sequence lengths. For MIND, we followed the preprocessing steps in \cite{recforest}, retaining interactions with at least 15 users and capping sequence lengths at 70. For Amazon-Office, we filtered out users with fewer than 5 interactions. To investigate the impact of item quantity, we split the Office dataset into Office (S) and Office (L), containing 14,000 and 37,000 items, respectively. We extracted categories, sub-categories, and titles from the MIND dataset, while utilizing type, brand, title, and category from the Amazon dataset as textual fields. We used Sentence-T5 \cite{sentenceT5} to obtain 768-dimensional semantic embeddings for each item.

\noindent{\bfseries Hyperparameter Settings.} 
The RQ-VAE model consists of three core components: an MLP encoder, a residual quantizer, and an MLP decoder. The encoder includes three hidden layers with dimensions [512, 256, 128], utilizing ReLU activations, and outputs a 96-dimensional latent representation. Each item is assigned a unique 3-tuple semantic token, with shared codebooks across the three levels, and the codebook size is set to 64. The hyperparameters are set as $\alpha=0.1$, $\beta=0.25$ and $\tau=0.1$. We utilize the Adam optimizer with a learning rate of 0.0001 and a batch size of 256 for codebook learning. For seq2seq transformer model training, we adopt the same structure as TIGER \cite{tiger}, setting the batch size to 512 and the learning rate to 0.001.



\noindent{\bfseries Evaluation Metrics.} We employ the widely-used metrics, Recall@$k$ and NDCG@$k$, to evaluate the effectiveness of our proposed method. Following the standard leave-one-out evaluation protocol, we reserve the last item for testing, the penultimate item for validation, and use the remaining items for training.
\begin{table}[]
\caption{Experimental results on the MIND dataset.}
\label{tab:MIND}
\begin{tabular}{c|c|cccc}
\toprule[1pt]
Methods & Metrics & @5 & @10 & @20 & @40 \\ 
\midrule
\multirow{2}{*}{Random} 
    & NDCG & 0.0201 & 0.0265 & 0.0327 & 0.0390 \\
    & Recall & 0.0319 & 0.0519 & 0.0766 & 0.1075 \\
\midrule
\multirow{2}{*}{$\mathcal{L}_{re}$} 
    & NDCG & 0.0363 & 0.0474 & 0.0594 & 0.0727 \\
    & Recall   & 0.0560 & 0.0905 & 0.1384 & 0.2031 \\
\midrule
\multirow{2}{*}{$\mathcal{L}_{co}$} 
    & NDCG & \textbf{0.0522} & \textbf{0.0663} & \textbf{0.0817} & \textbf{0.0975} \\
    & Recall & \textbf{0.0803} & \textbf{0.1241} & \textbf{0.1855} & \textbf{0.2625} \\
\midrule    
\multirow{2}{*}{$\mathcal{L}_{re}+\mathcal{L}_{co}$} 
    & NDCG & 0.0444 & 0.0574 & 0.0710 & 0.0865\\
    & Recall & 0.0677 & 0.1081 & 0.1621 & 0.2376\\
\midrule
\multirow{2}{*}{Impr.} 
    & NDCG & \textbf{43.76$\%$} & \textbf{39.90$\%$} & \textbf{37.52$\%$} & \textbf{34.10$\%$} \\
    & Recall & \textbf{43.34$\%$} & \textbf{37.21$\%$} & \textbf{34.05$\%$} & \textbf{29.24$\%$} \\
\bottomrule[1pt]
\end{tabular}
\vspace{-1ex}
\end{table}


\begin{table}[]
\caption{Experimental results on the Office dataset.}
\label{tab:office}
\begin{tabular}{c|c|cc|cc}
\toprule[1pt]
\multirow{2}{*}{Methods}  & \multirow{2}{*}{Metrics} & \multicolumn{2}{c|}{Office (S)} & \multicolumn{2}{c}{Office (L)} \\
\cmidrule(lr){3-4} \cmidrule(lr){5-6}
 &  & @10 & @20 & @10 & @20 \\ 
\midrule
\multirow{2}{*}{$\mathcal{L}_{re}$} 
& NDCG & 0.0024 & 0.0032 & 0.0041 & 0.0053 \\ 
& Recall & 0.0037 & 0.0068 & 0.0075 & 0.0123 \\
\midrule
\multirow{2}{*}{$\mathcal{L}_{co}$} 
& NDCG & 0.0034 & \textbf{0.0043} & \textbf{0.0047} & \textbf{0.0059} \\
& Recall & 0.0060 & 0.0094 &\textbf{0.0079} & \textbf{0.0125} \\
\midrule
\multirow{2}{*}{$\mathcal{L}_{re}+\mathcal{L}_{co}$} 
& NDCG & \textbf{0.0035} & {0.0042} & 0.0042 & 0.0052 \\
& Recall & \textbf{0.0066} & \textbf{0.0096} & \textbf{0.0079} & 0.0119 \\
\midrule
\multirow{2}{*}{Impr.} 
& NDCG & \textbf{43.80$\%$} & \textbf{31.06$\%$} & \textbf{15.11$\%$} & \textbf{10.53$\%$} \\ 
& Recall & \textbf{80.95$\%$} & \textbf{41.03$\%$} & \textbf{5.38$\%$} & \textbf{1.46$\%$} \\
\bottomrule[1pt]
\end{tabular}
\end{table}

\input{Figure/temperature.tex}

\subsection{Performance Comparison and Analysis}
\noindent{\bfseries Performance Comparison.} 
To validate our CoST method for semantic tokenization, we conducted experiments across various settings, including using random hashing and $\mathcal{L}_\text{re}$ as baselines, employing only $\mathcal{L}_\text{co}$, and combining  $\mathcal{L}_\text{re}$ with $\mathcal{L}_\text{co}$. Table \ref{tab:MIND} demonstrates that, on the MIND dataset, our CoST tokenization method outperforms these settings, showing significant improvements in both NDCG@k and Recall@k by approximately 38\% and 36\%, respectively. Table \ref{tab:office} presents the results for the Office (S) and Office (L) datasets. For shorter sequences and fewer samples, the combination of $\mathcal{L}_\text{co}+\mathcal{L}_\text{re}$ slightly outperforms $\mathcal{L}_\text{co}$ alone, whereas for longer sequences and more samples, $\mathcal{L}_\text{co}$ excels. This underscores the effectiveness of our CoST approach in the context of generative recommendation.


\noindent{\bfseries Sensitivity Analysis of Temperature $\tau$ and Training Epochs.}
To analyze how the temperature $\tau$ affects the training of CoST, we conducted experiments on the MIND dataset with  $\tau$ set to $\{0.1, 0.5, 1.0\}$. CoST was trained for 20 and 100 epochs, respectively, yielding the best Recall@40 when $\tau=1.0$ and $\tau=0.1$, as shown in Figure \ref{fig:Temperature}. This indicates that the performance of CoST varies with $\tau$ within a specific  range. Additionally, fixing $\tau=0.1$, we trained the model for 20, 50, and 100 epochs. As illustrated in Figure \ref{fig:Temperature}, with an increase in the number of training epochs, the loss steadily decreases, and the NDCG and Recall metrics consistently improve, demonstrating the effectiveness of CoST in training generative recommendation models.

\input{Figure/ablation}


\noindent{\bfseries Sensitivity Analysis of Codebooks.}
The effectiveness of semantic tokenization can be influenced by key hyperparameters, such as the codebook size ($K$), the number of codebooks ($M$), and the embedding dimension ($d$) of code vectors. As illustrated in Figure~\ref{fig:codebook}, as the codebook size and the number of codebooks increases, NDCG shows a consistent upward trend. This suggests that a larger token space allows for more effective representation of each item, leading to more precise tokenization. Additionally, as the embedding dimension increases from 32 to 96, the model's capability to represent items improves, leading to enhancements across various metrics. However, increasing the dimension to 256 results in a slight performance degradation due to overfitting.

%% file: Figure/temperature.tex
\begin{figure}[!t]

\begin{tabular}{cc}
\resizebox{.49\linewidth}{!}{
\includegraphics[width=\linewidth]{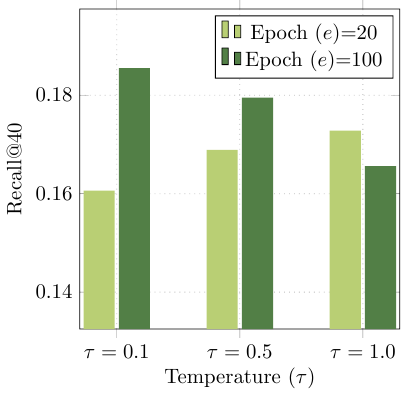}
}
\resizebox{.49\linewidth}{!}{
\includegraphics[width=\linewidth]{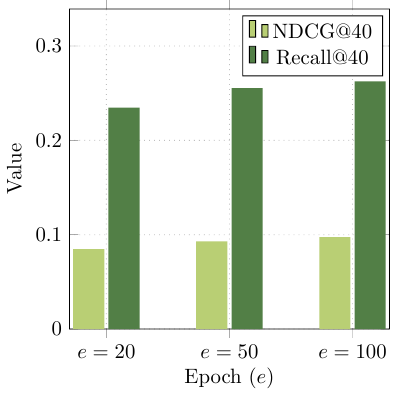}
}
\end{tabular}
\vspace{-2ex}
\caption{Analysis on temperature $\tau$ and training epochs $e$ on the MIND dataset.}
\label{fig:Temperature}
\vspace{-2ex}
\end{figure}

%% file: Figure/ablation.tex
\begin{figure}[!h]

\includegraphics[width=0.55\linewidth]{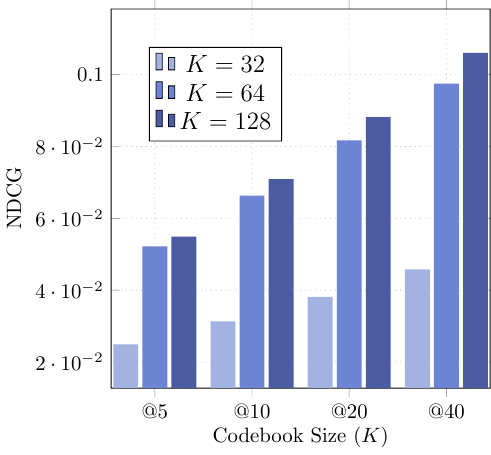}\vspace{1ex}

\includegraphics[width=0.55\linewidth]{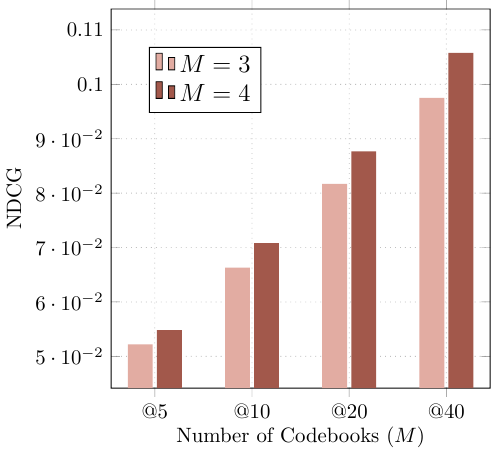}\vspace{1ex}

\includegraphics[width=0.55\linewidth]{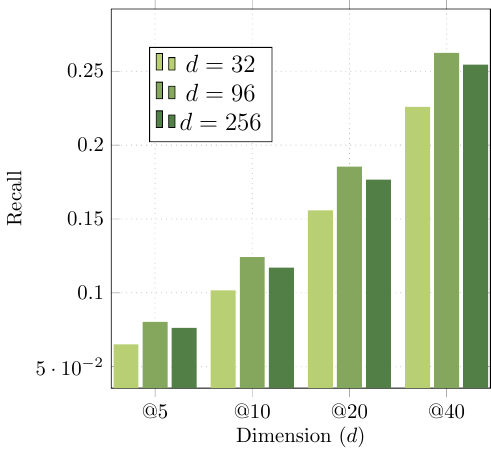}
\vspace{-1ex}
\caption{Sensitivity analysis on codebook size $K$ (fixed $M=3$ and $d=96$), number of codebooks $M$ (fixed $K=64$ and $d=96$), embedding dimension $d$ (fixed $K=64$ and $M=3$) on the MIND dataset.}
\label{fig:codebook}
\vspace{-2ex}
\end{figure}

%% file: Content/Conclusion.tex
\section{conclusion}
In this work, we delve deeper into semantic tokenization as a crucial preliminary step for generative recommendation tasks. To overcome the limitations of reconstruction quantization-based RQ-VAE, we introduce contrastive quantization into the tokenization process, resulting in the CoST method. By leveraging contrastive loss, which maximizes the top-one probability within a batch, CoST more effectively captures the underlying distribution of item similarities, providing a key advantage for recommendation tasks. Our experimental results demonstrate the effectiveness of CoST in enhancing generative recommendation performance. While CoST currently focuses on top-one neighborhood alignment, future work could extend this to consider top-k nearest neighbors. Looking ahead, we anticipate that further advancements in semantic tokenization, potentially incorporating multimodal information or user behavior data, will continue to improve the efficacy of generative recommender systems.

\begin{acks}
We gratefully acknowledge the partial support of MindSpore (\url{https://www.mindspore.cn}) for this work, which is a new deep learning computing framework.
\end{acks}